\documentclass[useAMS,usenatbib,referee]{mn2e}

\usepackage{graphicx}
\newcommand{\actaa}{Acta Astron.}
\newcommand{\apj}{ApJ}
\newcommand{\apjl}{ApJL}
\newcommand{\apjs}{ApJS}
\newcommand{\mnras}{MNRAS}
\newcommand{\pasp}{PASP}

\title[Microlensing event rates]
{Empirical microlensing event rates predicted by a phenomenological model}
\author[R. Poleski]
{Rados\l{}aw Poleski$^{1}$\thanks{E-mail: poleski.1@osu.edu}
\\
$^{1}$Department of Astronomy, Ohio State University, 140 West 18th Avenue, Columbus, OH 43210, USA\\
}
\begin{document}
\date{Accepted ???. Received ???; in original form ???}

\pagerange{\pageref{firstpage}--\pageref{lastpage}} \pubyear{2015}

\maketitle

\label{firstpage}

\begin{abstract}
Estimating the number of microlensing events observed in different parts of the Galactic bulge is a crucial point 
in planning microlensing experiments. 
Reliable estimates are especially important if observing resources are scarce, as is the case for space missions: K2, WFIRST, and Euclid. 
Here we show that the number of detected events can be reliably estimated based on statistics of stars observed in targeted fields. 
The statistics can be estimated relatively easily, which makes presented method suitable for planning future microlensing experiments.
\end{abstract}


\section{Introduction} 

The scientific yield of any gravitational microlensing experiment strongly depends on the number of detected events. 
A higher number of events improves the statistical properties of the event sample and allows one to detect larger number of microlensing anomalies, 
e.g., planetary signals provided that the observing cadence is high enough. 
Hence, it is of primary importance to predict how many events would be detected in a given part of the Galactic bulge. 
Additionally, the event rate gives us information about the structure of the bulge \citep[e.g.,][]{kiraga94}.  
The connection between statistics of observed events and the structure of the bulge is an area of active research \citep{kerins09,sumi13,wyrzykowski15}, 
but a full understanding of microlensing event rate has not yet been achieved. 

Here we present a new approach to a problem of predicting microlensing event rates. 
We show that the number of events observed scales with the product of the surface density of stars brighter than the completeness limit ($N_{*}(I < I_{\rm lim})$) 
and the surface density of red clump (RC) stars in given field ($N_{\rm RC}$). 
The RC stars are relatively bright and easy to identify. 
Hence, the RC sample is essentially complete in any given field, except the highest extinction fields. 
Thanks to their brightness, number of RC stars is a good proxy for total number of stars observed in given sight-line i.e., 
they estimate the number of potential lenses in given field. 
The surface density $N_{*}(I < I_{\rm lim})$ is the  proxy for a number of potential sources in microlensing events. 
Hence, one may conjecture that the number of events observed is proportional to the product of number of potential lenses and number of potential sources. 
This is obviously true if the stellar kinematics and the distribution of stars along the line of sight did not changed from field to field. 
It is important to note that both $N_{*}(I < I_{\rm lim})$ and $N_{\rm RC}$ can be relatively easily measured. 
The next section summarizes the data used. 
In Section 3 we present the basic correlation we found. 
Section 4 describes efforts to find tighter relation. 
We discuss results in Section 5.

\section{Data} 

We analyse the catalogue of standard microlensing events presented by \citet{wyrzykowski15}. 
The events are standard in the sense that they do not show obvious signatures of either binary lenses or sources, nor extended source effects. 
Microlensing parallax effect may be pronounced in some of these events 
but they are considered standard as long as the light curve is well fitted by a point-source point-lens model. 
We prefer a catalogue of standard events selected after the observations are completed over the list of candidate events selected while the observations are ongoing, because the former is more objective, more uniform, more complete, and contains smaller number of false-positives. 
\citet{wyrzykowski15} searched photometric databases of 
the third phase of the Optical Gravitational Lensing Experiment \citep[OGLE-III; ][]{udalski03,udalski08red}. 
The sky images were collected between June 2001 and May 2009 using the 1.3-m telescope at Las Campanas Observatory, Chile 
and an eight chip camera with a field of view of $36^\prime\times36^\prime$. 
Microlensing events were searched after the project was completed, thus events peaking during whole project time span were found. 
We note that $15~{\rm per cent}$ of sources are RC stars. 
Different astrophysical phenomena that can mimic microlensing events, like dwarf nova outbursts or brightening of Be stars, 
were rejected thanks to the long term photometry. 

\begin{figure}
\includegraphics[bb=20 158 563 463,width=\columnwidth]{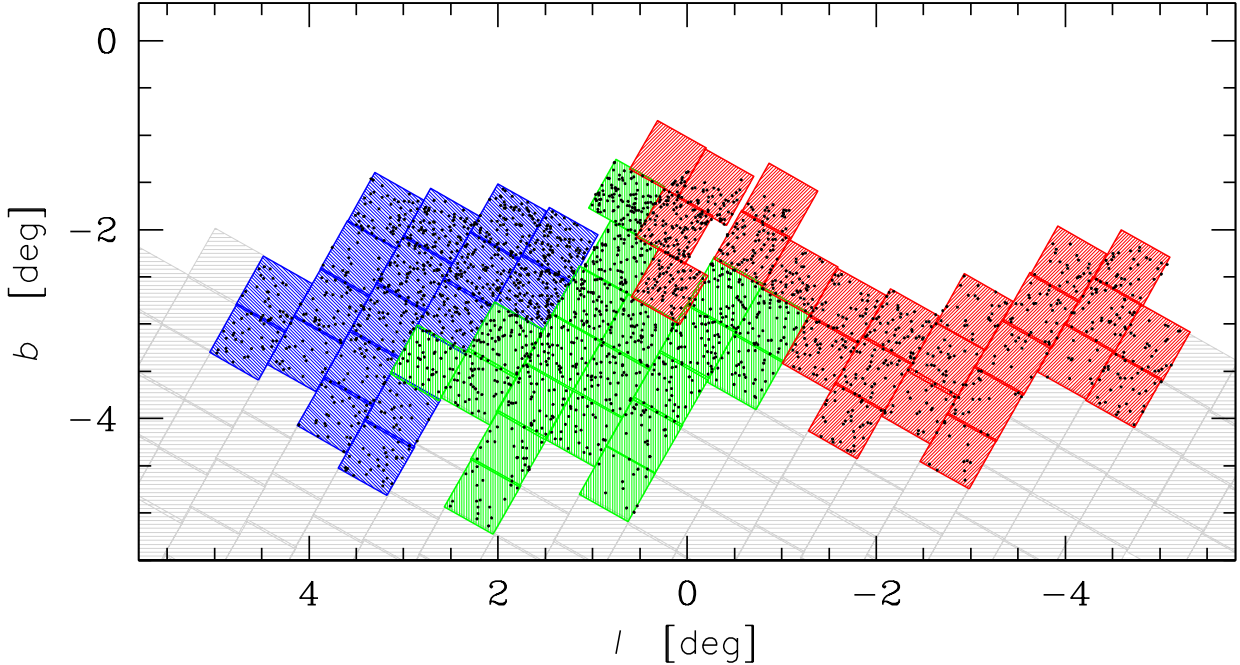}
\caption{Analysed OGLE-III fields (color-shaded squares) and microlensing events (black points) in Galactic coordinates. 
Fields are divided into three groups marked by different colors as described in Sec.~\ref{sec:disc}. 
Gray shaded fields are not taken into account because of too small number of epochs. 
Three red-shaded fields have subfields with extinction that is too high to measure RC stars and hence we omit them in present study. 
\label{fig:fields}}
\end{figure}

There are a variety of  factors that affect the number of detected microlensing events, 
some of which are of astrophysical origin and other are survey-specific biases. 
An important factor, that we are not able to fully eliminate, is different observing cadence in different fields.  
Additionally, the number of epochs in a given field changes from one year to another. 
Differing observing cadence affects the efficiency with which surveys can find events, especially short events. 
We minimize the effect of changing cadence by selecting for our analysis a subset of fields and a subset of observing seasons. 
We remove events peaking in 2001, as the OGLE-III survey started in the middle of the 2001 bulge season and the observing strategy differed significantly from other years. 
The last year of the OGLE-III observations ended before the best observing conditions for bulge fields started in June. 
Thus, we also eliminate events peaking in 2009. 
Other than that, the biggest change in observing strategy was introduced in 2004 when many fields were abandoned 
and the observations focused on fields with higher number of events; hence, we decided to analyse only events peaking between 2004 and 2008. 
Even after imposing these criteria we are left with fields observed at different cadence. 
In order to impose reasonable uniformity we excluded fields with less than 90 epochs per year. 
This leaves us with 63 fields for which mean number of epochs per year varies between 94 and 262 with average of 165 and a total of 2563 events. 
The centres of these fields are within Galactic coordinates range $-4\fdg9 < l < 4\fdg6$ and $-4\fdg8 < b < -1\fdg3$ 
and are outlined in Fig.~\ref{fig:fields}.

In addition to the information about events, we also need data that describe the structure of the Galaxy. 
We employ two kinds of data: the surface density of RC stars and the surface density of all stars down to a certain limit. 
The RC stars in the OGLE-III bulge fields were investigated in detail by \citet{nataf13b} and we use their results. 
\citet{nataf13b} divided the OGLE-III fields into a number of sight lines 
and provide spatial density of RC stars for each sight line separately. 
In the area considered here, three fields have 
subfields\footnote{Subfield is sky area corresponding to single CCD chip i.e., 1/8 of camera field of view.} 
for which \citet{nataf13b} could not measure the properties of RC because of extremely high extinction. 
These sky-areas are not used in our study. 
We divided every subfield into 32 regions and interpolated RC spatial density values to the centres of the regions. 
To find the $N_{\rm RC}$ for each subfield we averaged the interpolated values. 
In this way we cope with varying density of \citet{nataf13b} sightlines and edge effects. 

The last piece of information we need is spatial density of stars observed in each sky area. 
We use the photometric catalogues presented by \citet{szymanski11} that list stars detected on the OGLE-III reference images. 
In each subfield we count the number of stars down to specified brightness limit and scale it to $1.0~{\rm deg^2}$. 
Note, that each reference image analysed by \citet{szymanski11} was constructed using a set of high-quality images, 
but slightly shifted with respect to a defined centre. 
This renders the outer parts of the subfields of lower photometric quality 
and hence they are not considered for estimating spatial density of stars.

\section{Primary correlation} 

\begin{figure}
\includegraphics[width=\columnwidth]{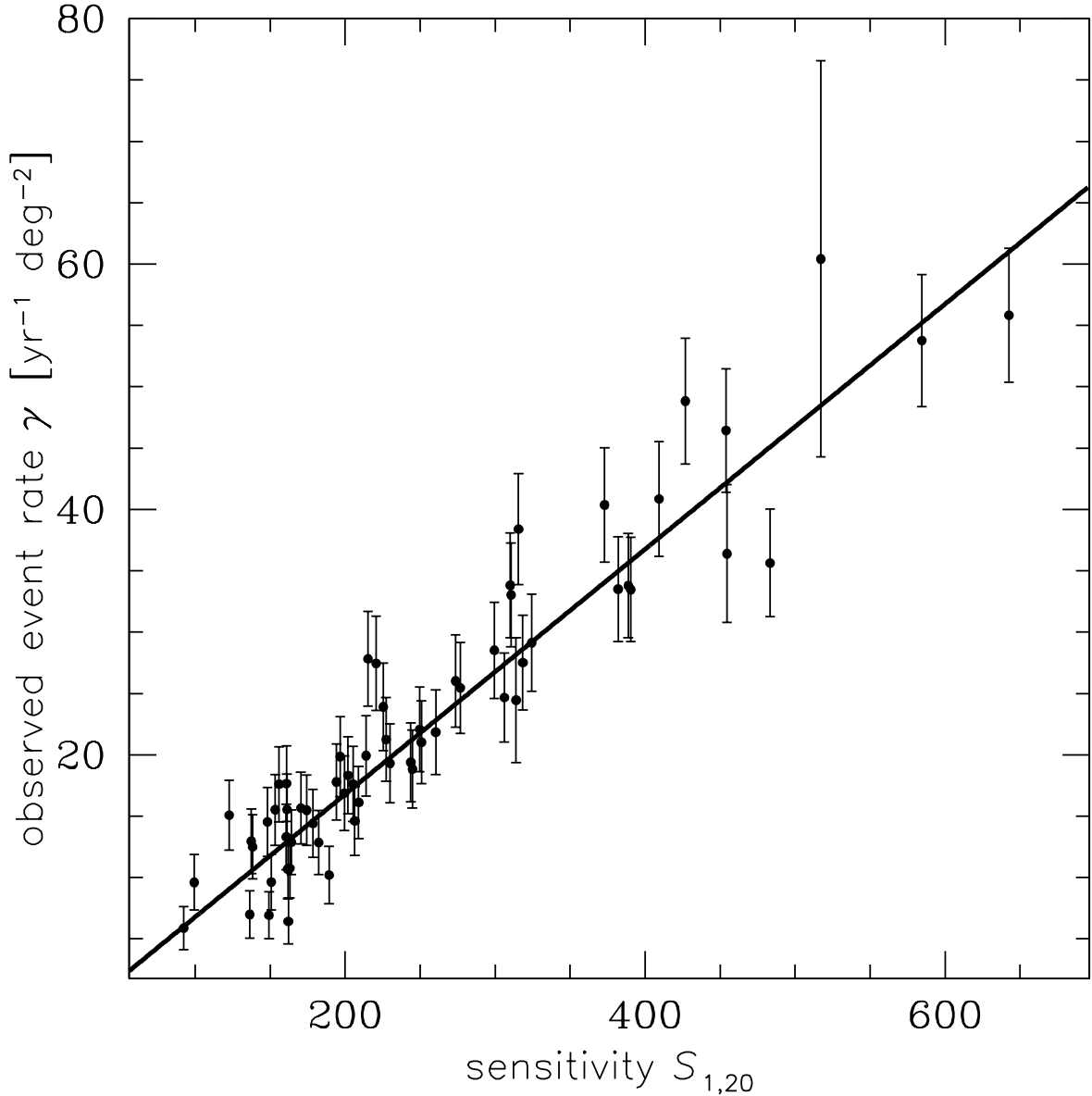} 
\caption{Relation between number of observed microlensing events in a given field and product of surface densities of stars brighter than $20~{\rm mag}$ and RC stars. 
The single point with large errorbar represents field blg194 in which only a single subfield has $N_{\rm RC}$ derived. 
\label{fig:firstcorr}}
\end{figure}

For each field we derive mean number of events per year per ${\rm deg^2}$ 
($\gamma$; this symbol is used in analogy to frequently used $\Gamma$ as an event rate corrected for the detection efficiency). 
We define microlensing event sensitivity as:
\begin{equation}
S_{\alpha, I_{\rm lim}} \equiv \left(\frac{N_{\rm RC}}{10^3~{\rm deg^{-2}}}\right)^{\alpha} \times \left(\frac{N_{*}(I < I_{\rm lim})}{10^6~{\rm deg^{-2}}}\right).
\label{equ:prim}
\end{equation}
We start with a simple choice of $\alpha = 1$ and $I_{\rm lim} = 20~{\rm mag}$. 
This limiting brightness is chosen because it is a rough estimate of the OGLE-III completeness in the dense stellar fields. 
For reference we provide the average numbers for the well-studied Baade window ($l=1\fdg0,~b=-3\fdg9$): 
$N_{\rm RC} = 49.9$, 
$N_{*}(I < 20\,{\rm mag}) = 4.05$, and 
$N_{*}(I < 20.5\,{\rm mag}) = 4.68$ 
in the same units as in Eq.~\ref{equ:prim}. 
We present plot of $\gamma$ as a function of $S_{1, 20}$ in Fig.~\ref{fig:firstcorr}. 
The line fitted to these data is:
\begin{equation}
\frac{\gamma}{{\rm deg^{-2} yr^{-1}}} = 0.0999(44) \times S_{1, 20} - 3.21(97)
\label{equ:firstcorr}
\end{equation}
with $\chi^2/{\rm dof} = 1.46$. 
The uncertainties in $S_{1, 20}$ are negligible compared to the Poisson noise of $\gamma$. 
It turns out that even our simple approach for finding sensitivity results in a relatively tight correlation. 

\section{Detail parameter selection} 

There are multiple effects that may affect the number of potential lenses e.g., the number of disk to bulge stars changes with Galactic latitude.  
Yet, our definition of sensitivity takes only $N_{\rm RC}$ into account. 
Hence we allowed $\alpha$ to vary and selected the value that gives the tightest relation for given $I_{\rm lim}$. 

We further tried to improve our event rate estimates by revising microlesning events included. 
In the fields with smaller number of epochs the shortest events are strongly affected by selection biases, thus, 
we checked the impact of rejecting all events with Einstein time-scale shorter than some limit $t_{\rm E,lim}$ 
(set to between $5~{\rm d}$ and $10~{\rm d}$). 
The reduction of the scatter in the $\gamma(t_{\rm E} > t_{\rm E,lim})$-$S_{\alpha,I_{\rm lim}}$ relation is only slight. 
In our final calculation we used $t_{\rm E,lim} = 8~{\rm d}$, 
which resulted in removing $11~{\rm per~cent}$ of the event sample. 
In eighteen best-observed fields (at least 1200 epochs in OGLE-III) $14~{\rm per~cent}$ of the sample was removed. 

\begin{figure}
\includegraphics[width=\columnwidth]{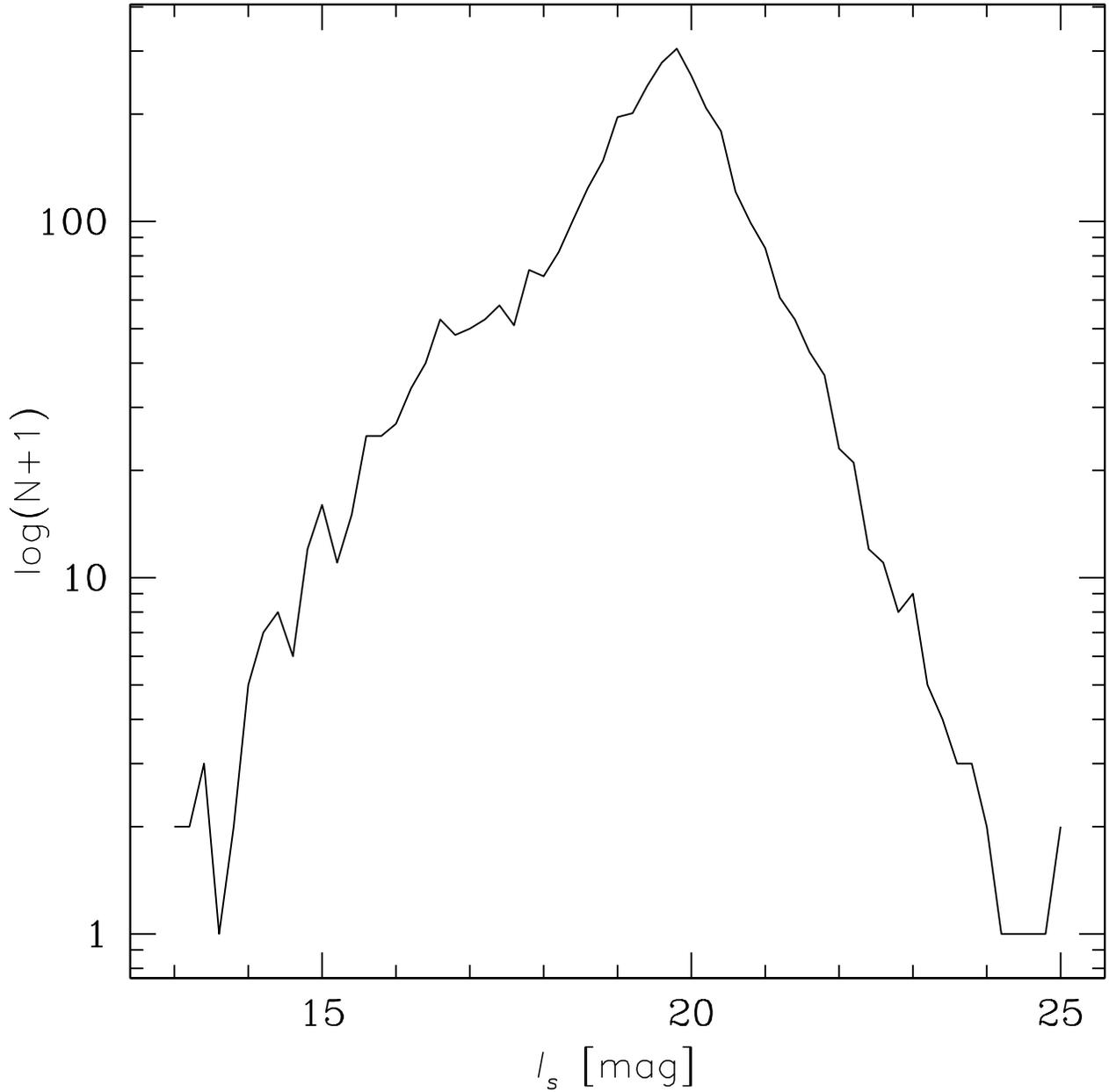}
\caption{Histogram of source brightness in \citet{wyrzykowski15} sample. The bins are $0.2~{\rm mag}$ wide.}
\label{fig:Ishist}
\end{figure}

An important simplification we made in previous section is that our proxy for number of potential sources 
is the number of stars down to a sharp brightness limit $I_{\rm lim} = 20~{\rm mag}$. 
In a real experiment, no such sharp limit exists; 
sources can be significantly fainter than the photometric detection limit and still be detected if they are highly magnified. 
On the other hand, almost all events with bright sources are easily found. 
Instead of detailed modelling of detection efficiency that could take into account the event time-scales, 
observing cadence, luminosity function etc., 
we try to improve our relation while retaining its simplicity. 
We calculate the source brightness ($I_s$) for events presented by \citet{wyrzykowski15} based on their blending ratios and baseline brightness. 
In Fig~\ref{fig:Ishist} we show a histogram of $I_s$. 
The number of detected events falls to half of maximum at $I = 20.5~{\rm mag}$, 
and this limit is used for further calculations of number of stars. 

\begin{figure}
\includegraphics[angle=270,width=\columnwidth]{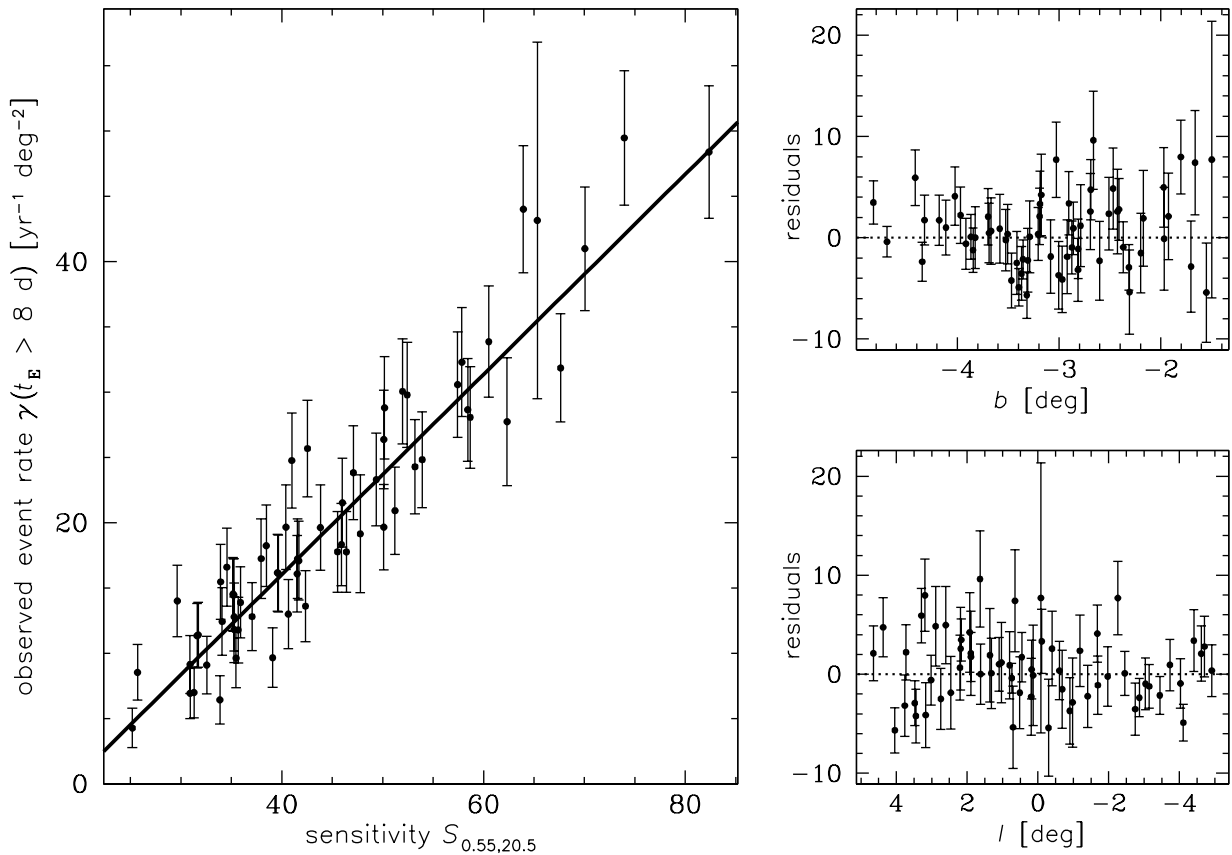}
\caption{Final relation between number of observed microlensing events with $t_{\rm E} > 8~{\rm d}$ in given field and $S_{0.55,20.5}$ sensitivity.
Right panels present residuals as a function of Galactic coordinates.  
\label{fig:secondcorr}}
\end{figure}

Our final relation is presented in Fig.~\ref{fig:secondcorr} and fitted with: 
\begin{equation}
\frac{\gamma(t_{\rm E} > 8~{\rm d})}{{\rm deg^{-2} yr^{-1}}} = 0.767(37) \times S_{0.55, 20.5} - 14.6(15). 
\label{equ:secondcorr}
\end{equation}
The resulting $\chi^2/{\rm dof} = 1.17$  
and is significantly smaller than in primary relation. 
Fig.~\ref{fig:secondcorr} also presents how residuals from fitted relation depend on Galactic longitude and latitude. 
No significant dependence is seen. 
This shows robustness of our method. 

The two relations presented in Eq.~\ref{equ:firstcorr} and \ref{equ:secondcorr} 
show that the choice of $\alpha$ and $I_{\rm lim}$ may significantly impact how tight the relation is. 
In order to further verify how optimal value of $\alpha$ depends on $I_{\rm lim}$ 
we present a plot of $\chi^2/{\rm dof}$ as a function of $I_{\rm lim}$ and $\alpha$. 
One clearly sees that both parameters are anti-correlated and 
the best fits correspond to $I_{\rm lim}$ in a range from $19.9$ to $20.6~{\rm mag}$. 

\begin{figure}
\includegraphics[bb=30 160 598 633,width=\columnwidth]{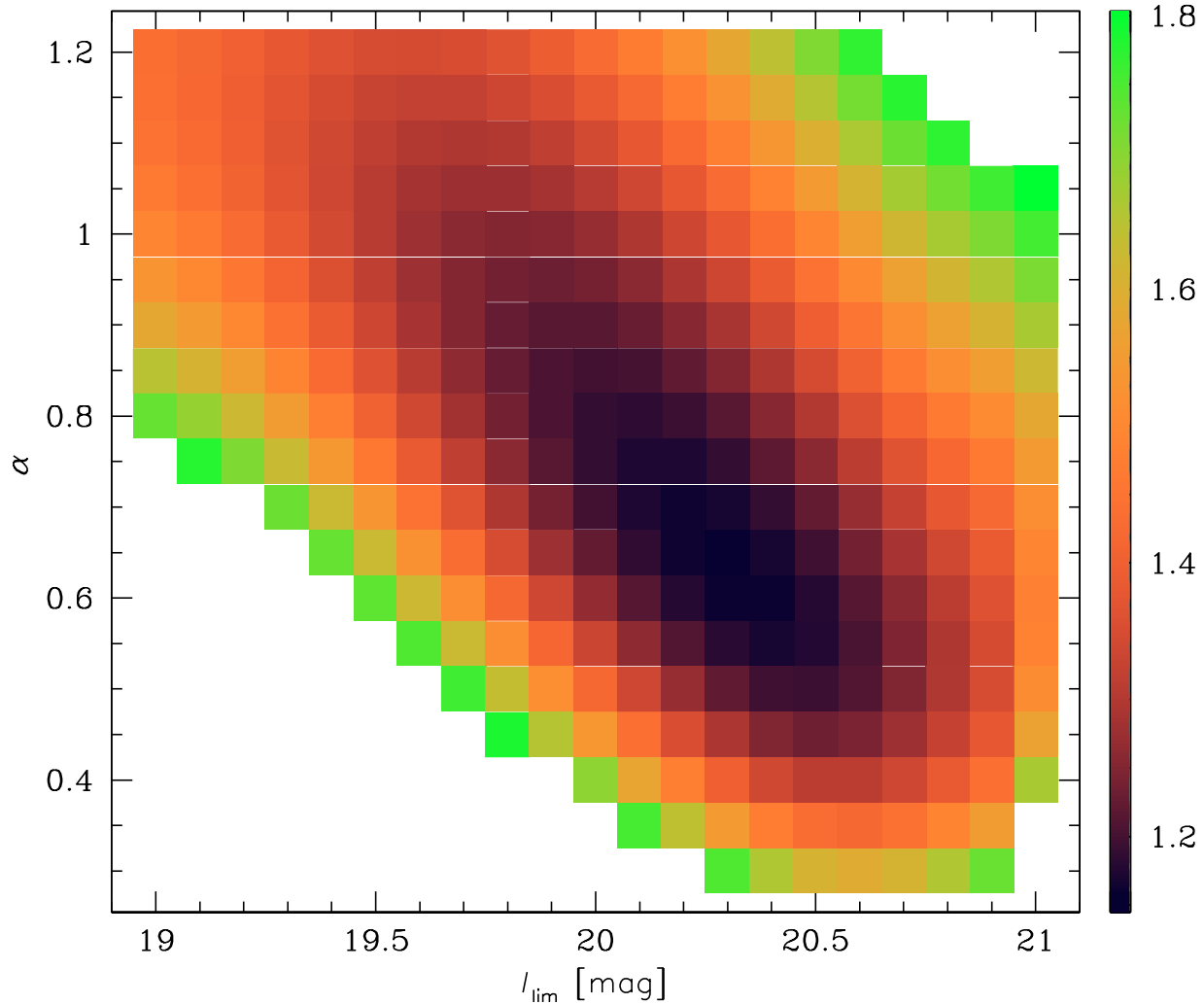}
\caption{The $\chi^2/{\rm dof}$ surface in $I_{\rm lim}-\alpha$ space. 
\label{fig:chi2}}
\end{figure}

\section{Discussion} 
\label{sec:disc}

We have shown that the number of observed microlensing events can be estimated based on the statistics of stars observed in a given field. 
It is important to note that presented framework takes into account all stellar populations that constitute the bulge and disk, 
including the ones that have not yet been fully described. 
As an example, the X-shaped structure was hidden for a long time and discovered only recently \citep{nataf10,mcwilliam10}. 
Our method takes such structures into account via $N_{*}(I < I_{\rm lim})$ and, 
unlike efforts that depend on the Galactic distribution models, 
does not depend on current knowledge of different stellar populations. 

We note that better understanding of different factors contributing to bulge microlensing could be achieved via a two-step investigation.  
First, the Galactic distribution models would give reliable predictions of 
the number density of brightness-limited sample of stars stars and the number density of RC stars. 
Second, one would use relations presented here to estimate the 
expected event rates and compare these to the measured values.  
The events that appear in the sky-areas with low sensitivity have higher chance of being disk-disk events. 
There is particular interest in events involving globular cluster lenses. 
Cluster lenses have known distances and proper motions, hence one can measure their mass based on $t_{\rm E}$ only \citep{paczynski94,pietrukowicz12}. 
Our framework could be used to select promising candidates for cluster lensing events: 
the events appearing in low sensitivity regions and near the central regions of a globular cluster. 

The key application of the relations presented above will be in selection of particular fields for satellite survey observations. 
The planning of survey fields requires comparison of expected yields on different angular scales. 
We tried to verify if our relations hold in different areas of Galactic bulge in two different ways. 
First, we divided analysed fields into three groups with following constrains: 
each group represents continuous (or near-continuous) sky-area, 
the three fields with the highest event rate are in different groups, and 
total number of events in each group is similar. 
The division of OGLE-III fields in the three groups is presented by different colours in \ref{fig:fields}. 
The groups differ mostly in mean Galactic longitude. 
The parameters of each group and the results of line fitting are presented in Table~\ref{tab:groups}. 
There are no significant differences in values of parameters fitted between different 
groups as well as between them and relation presented in Eq.~\ref{equ:secondcorr}. 
On the other hand, the $\chi^2/{\rm dof}$ varies by more than a factor of three. 
It is the largest in the East part of bulge. 

\begin{table}
 \begin{tabular}{l|r|r|r|r|r|r|r}
 Color in Fig~\ref{fig:fields} & $n_{\rm fields}$ & $n_{\rm events}$ & $<l>$ & $<b>$ & $a~[{\rm deg^{-2}yr^{-1}}]$ & $b~[{\rm deg^{-2}yr^{-1}}]$ & $\chi^2/{\rm dof}$ \\
\hline\hline
  blue  & $ 18 $ & $ 853 $ & $  3\fdg07 $ & $ -2\fdg86 $ & $ 0.792(76) $ & $ -15.3(33) $ & $ 2.11 $ \\
  green & $ 20 $ & $ 855 $ & $  0\fdg90 $ & $ -3\fdg45 $ & $ 0.734(56) $ & $ -12.7(23) $ & $ 0.62 $ \\
  red   & $ 25 $ & $ 855 $ & $ -2\fdg40 $ & $ -2\fdg92 $ & $ 0.774(65) $ & $ -15.6(26) $ & $ 1.00 $ \\
\hline
 \end{tabular}
 \caption{Comparison of event rate vs. sensitivity relation fits in groups of fields outlined in Fig~\ref{fig:fields} by different colours. 
  Columns give: colour of shading in Fig~\ref{fig:fields}, number of fields in a group, 
  total number of events, mean Galactic coordinates, and linear fits results in the last three columns. 
  The relations fitted were in the form: $\gamma(t_{\rm E} > 8 d) = a \times S_{0.55,20.5} + b$, 
  i.e., the same as in Eq.~\ref{equ:secondcorr} for whole sample. 
 }
 \label{tab:groups}
\end{table}

\begin{figure}
\includegraphics[width=\columnwidth]{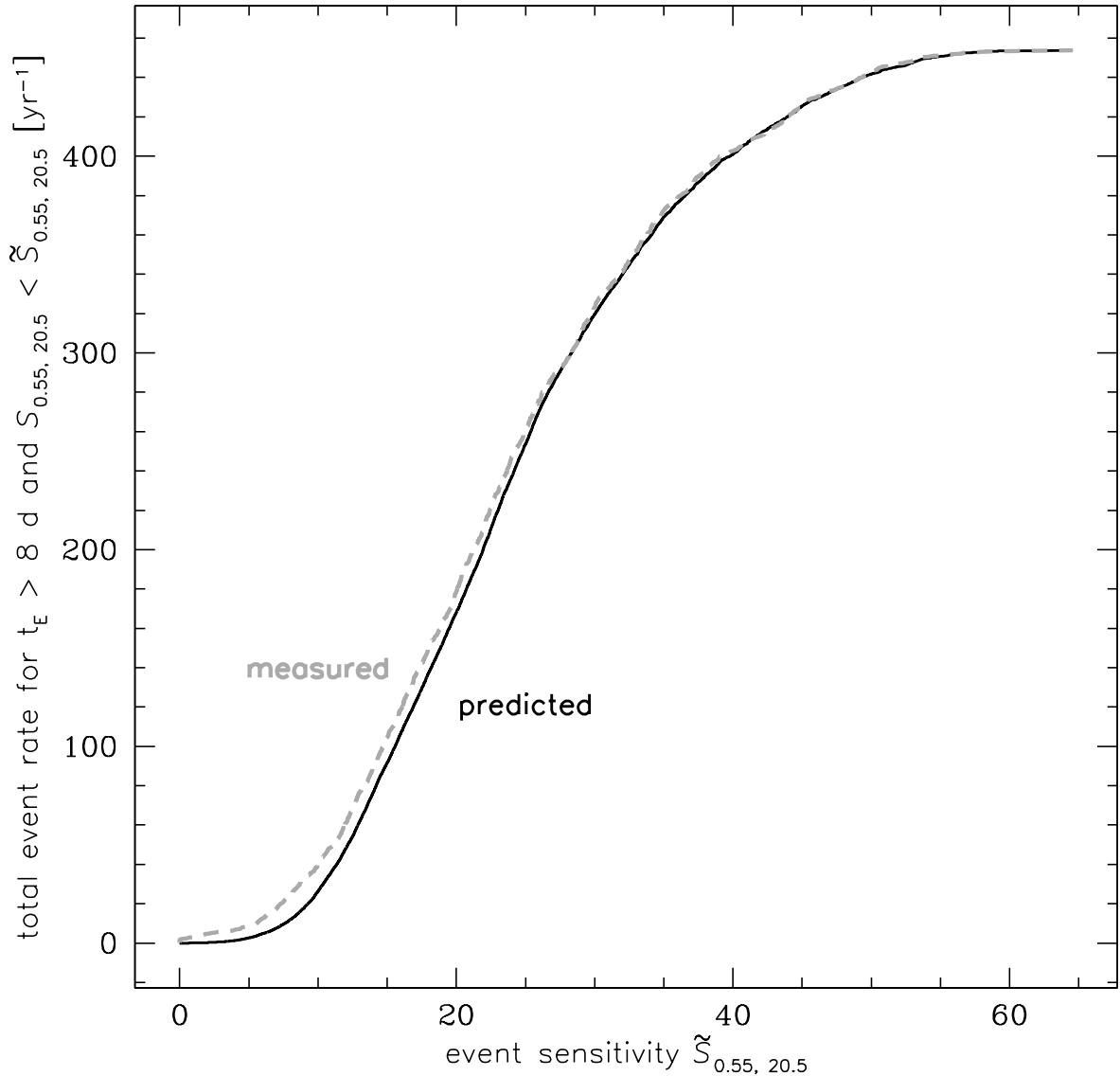}
\caption{Comparison of predicted and measured event rates for all $5~{\rm arcmin^2}$ 
sky-regions with sensitivity $S_{0.55, 20.5}$ smaller than given value $\tilde{S}_{0.55, 20.5}$. 
The predicted event rate is multiplied by $1.03$ to match the total observed event rate. 
This correction is smaller than the uncertainties in Eq.~\ref{equ:secondcorr}. 
\label{fig:smallS}}
\end{figure}

The above test verified how our empirical relation works on scales of a few degrees, 
but choosing optimal fields for microlensing observations requires also a comparison of sky-areas on arcminute scales. 
Thus, the second way we tested our relation is based on $5~{\rm arcmin^2}$ sky-regions used to estimate $N_{\rm RC}$. 
For each region we calculated sensitivity $S_{0.55, 20.5}$ and expected event rate. 
In Figure~\ref{fig:smallS} 
we compare the measured and expected event rate for all regions with sensitivity $S_{0.55, 20.5} < \tilde{S}_{0.55, 20.5}$. 
Note that to find expected event rate we integrate over small regions that are well separated on the sky. 
The measured and predicted values are very similar for the range of $\tilde{S}_{0.55, 20.5}$ used in deriving Eq.~\ref{equ:secondcorr}
i.e., for $\tilde{S}_{0.55, 20.5}>25$ the relative difference is smaller than $3\%$. 
On the other hand, for smaller $\tilde{S}_{0.55, 20.5}$ the differences are larger e.g., for $\tilde{S}_{0.55, 20.5}=12.5$ the relative difference is $21\%$. 
The overall similarity of the two curves in Fig.~\ref{fig:smallS} proves that our method reliably predicts event rate. 

\begin{figure}
\includegraphics[width=\columnwidth]{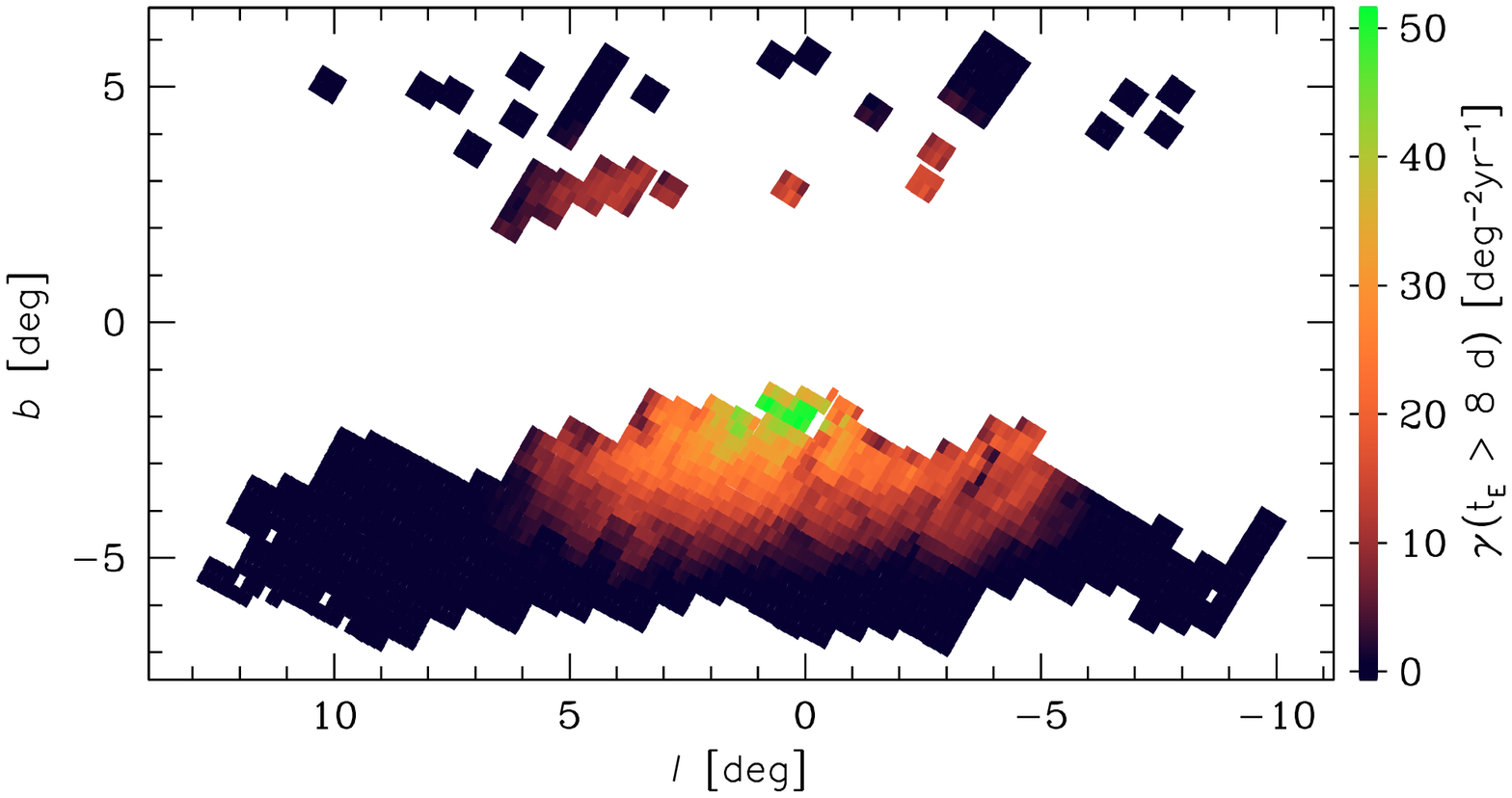}
\caption{Expected event rates for all bulge OGLE-III subfields with derived $N_{\rm RC}$. 
The values are based on Eq.~\ref{equ:secondcorr}.
\label{fig:mapS}}
\end{figure}

The empirical relation given in Eq.~\ref{equ:firstcorr} and \ref{equ:secondcorr} was derived 
based on a subset of OGLE-III bulge fields but can be used to predict 
the event rates in other fields as well. 
We estimate the values $N_{\rm RC}$ and $N_{*}(I < 20.5\,{\rm mag})$ in the same way as previously 
and average the expected event rate $\gamma(t_{\rm E} > 8~{\rm d})$ over the area of a subfield. 
For $2\%$ of the subfields the $N_{\rm RC}$ values are not available. 
The expected event rates are presented in Figure~\ref{fig:mapS}. 
The highest value of $\gamma(t_{\rm E} > 8~{\rm d})$ are close to $50\,{\rm deg^{-2}yr^{-1}}$ 
and are concentrated in small sky area $l\approx0\fdg5$ and $b\approx-2^{\circ}$.   
Most of the sky-area at $|l| > 6^{\circ}$ or $|b|>5^{\circ}$ shows very small event rates. 
All the estiamted values are provided in Table~\ref{tab:online1}.

\begin{table}
\begin{tabular}{l|r|r|r|r|r|r|r}
Name & $l$ & $b$ & $N_{\rm RC}$ & $N_{*}(I<20\,{\rm mag})$ & $ N_{*}(I<20.5\,{\rm mag})$ & $\gamma$ & $\gamma(t_{\rm E}>8\,{\rm d})$ \\
 & ${\rm [deg]}$ & ${\rm [deg]}$ & ${\rm [10^3\,deg^{-2}]}$ & ${\rm [10^6\,deg^{-2}]}$ & ${\rm [10^6\,deg^{-2}]}$ & ${\rm [deg^{-2}yr^{-1}]}$ & ${\rm [deg^{-2}yr^{-1}]}$ \\
\hline\hline
blg100.1 & $ -0.4198 $  & $ -1.7976 $  & $ 130.1 $  & $ 4.171 $  & $ 4.794 $  & $ 51.00 $  & $ 38.78 $  \\
blg100.2 & $ -0.2893 $  & $ -1.7199 $  & $ 125.2 $  & $ 3.840 $  & $ 4.708 $  & $ 44.69 $  & $ 36.64 $  \\
blg100.3 & $ -0.1591 $  & $ -1.6416 $  & $ 127.2 $  & $ 3.662 $  & $ 4.515 $  & $ 43.28 $  & $ 35.04 $  \\
blg100.4 & $ -0.0282 $  & $ -1.5635 $  & $ 129.7 $  & $ 3.470 $  & $ 4.334 $  & $ 41.90 $  & $ 33.65 $  \\
blg100.8 & $ -0.5732 $  & $ -1.5417 $  & $ 137.5 $  & $ 2.430 $  & $ 3.134 $  & $ 29.86 $  & $ 21.18 $  \\
blg101.1 & $ 0.0165 $  & $ -2.2121 $  & $ 107.4 $  & $ 5.606 $  & $ 6.192 $  & $ 56.97 $  & $ 47.49 $  \\
blg101.2 & $ 0.1464 $  & $ -2.1357 $  & $ 111.9 $  & $ 5.862 $  & $ 6.400 $  & $ 62.25 $  & $ 50.98 $  \\
blg101.3 & $ 0.2781 $  & $ -2.0575 $  & $ 116.3 $  & $ 5.716 $  & $ 6.107 $  & $ 63.32 $  & $ 49.37 $  \\
blg101.4 & $ 0.4084 $  & $ -1.9808 $  & $ 118.6 $  & $ 5.528 $  & $ 6.040 $  & $ 62.17 $  & $ 49.26 $  \\
blg101.5 & $ 0.2576 $  & $ -1.7233 $  & $ 128.5 $  & $ 4.591 $  & $ 5.099 $  & $ 55.79 $  & $ 41.82 $  \\
\ldots & & & & & & & \\
\hline
\end{tabular}
\caption{Expected event rates for OGLE-III bulge subfields. Mean surface densities of RC stars and all stars brighter brighter than given limit are also given.  
The last two columuns give expected event rates based on Eq.~\ref{equ:firstcorr} and \ref{equ:secondcorr}. 
The full version of the table containing 2084 records is available in the online version of the paper. 
}
\label{tab:online1}
\end{table}

Our empirical relations are based on a subsample of fields and events observed by the OGLE-III survey. 
Similar relations can be found for other high-cadence bulge observing surveys if corrected for different cadence, band-pass etc. 
The main application of such efforts should be selection of fields for microlensing surveys. 
We note that field selection does not depend on the zero point of scaling relation; 
it is enough to find relative number of events that a given survey should find in different fields to select optimal fields for observing. 
The highest priority i.e., space-based surveys, are K2 \citep{gould13b,howell14}, WFIRST \citep{spergel15}, and Euclid \citep{penny13}. 
Our method could also be applied in the process of field selection for LSST, if the Galactic disk fields are observed frequently enough to allow detection of microlensing events \citep{gould13c}.

\section*{Acknowledgments}

I deeply thank Andrew Gould for insightful discussions. 
The anonymous referee comments that help to improve the paper are acknowledged. 

\bibliographystyle{mn2e}

\bsp

\label{lastpage}

\end{document}